\documentclass[prd,preprint,superscriptaddress,showpacs,nofootinbib,%
tightenlines ]{revtex4}
\usepackage{graphics}
\usepackage{epsfig}

\newcommand{\logM}{\ln\frac{M}{\mu}}
\newcommand{\logMsq}{\ln^2\frac{M}{\mu}}

\begin{document}
\preprint{MKPH-T-06-19}
\title{Chiral expansion of the nucleon mass to order ${\cal O}(q^6)$}
\author{Matthias R.~Schindler}
\affiliation{Institut f\"ur Kernphysik, Johannes
Gutenberg-Universit\"at, 55099 Mainz, Germany}
\author{Dalibor Djukanovic}
\affiliation{Institut f\"ur Kernphysik, Johannes
Gutenberg-Universit\"at, 55099 Mainz, Germany}
\author{Jambul Gegelia}
\affiliation{Institut f\"ur Kernphysik, Johannes Gutenberg-Universit\"at, 55099
Mainz, Germany} \affiliation{High Energy Physics Institute, Tbilisi State University,
Tbilisi, Georgia}
\author{Stefan Scherer}
\affiliation{Institut f\"ur Kernphysik, Johannes Gutenberg-Universit\"at, 55099
Mainz, Germany}
\begin{abstract}
   We present the results of a complete two-loop calculation at order
${\cal O}(q^6)$ of the nucleon mass in manifestly
Lorentz-invariant chiral perturbation theory. The renormalization
is performed using the reformulated infrared renormalization,
which allows for the treatment of two-loop integrals while
preserving all relevant symmetries, in particular chiral symmetry.
\end{abstract}
\date{December 13, 2006}
\maketitle

\section{Introduction}
   Chiral perturbation theory (ChPT)
\cite{Weinberg:1978kz,Gasser:1983yg,Gasser:1984gg} is the
effective field theory for the strong interactions at low energies
and has been highly successful in the application to purely
mesonic processes (for a review see, e.g.,
Ref.~\cite{Scherer:2002tk}).
   The extension to the one-nucleon sector was first addressed in
Ref.~\cite{Gasser:1987rb} and originally seemed to be problematic
due to the presence of the nucleon mass as an additional mass
scale that does not vanish in the chiral limit.
   When using dimensional regularization in combination with
the modified minimal subtraction scheme there is no direct
correspondence between the loop expansion and the chiral
expansion.
   The first solution to this power counting problem was given by
heavy-baryon ChPT (HBChPT) \cite{Jenkins:1990jv,Bernard:1992qa},
in which an expansion in inverse powers of the nucleon mass is
performed in the Lagrangian.
   Later, several manifestly Lorentz-invariant renormalization schemes
have been developed that also result in a proper power counting
\cite{Ellis:1997kc,Becher:1999he,Gegelia:1999gf,Gegelia:1999qt,Goity:2001ny,Fuchs:2003qc},
   with the infrared (IR) regularization of
Ref.~\cite{Becher:1999he} the most commonly used scheme.
   In its original formulation IR regularization is applied to
one-loop intgrals, while the reformulated version of
Ref.~\cite{Schindler:2003xv} is also applicable to multi-loop
diagrams \cite{Schindler:2003je}.
   A different generalization of IR regularization to
two-loop diagrams was suggested in Ref.~\cite{Lehmann:2001xm}.

   Calculations at the two-loop level can be used to test
the convergence behavior of the chiral expansion, which, compared
to the purely mesonic part of the theory, seems to be slower in
the baryonic sector.
   This is also of interest to lattice calculations, where chiral
extrapolations are performed to obtain quantities at physical pion
masses (see, e.g., Ref.~\cite{Meissner:2005ba}).
   The nucleon mass at order ${\cal O}(q^5)$
has been analyzed in the framework of HBChPT
\cite{McGovern:1998tm}, including the evaluation of two-loop
diagrams.
   In Ref.~\cite{Bernard:2006te} renormalization group techniques were
used to determine the leading nonanalytic contributions to the
nucleon axial-vector coupling constant $g_A$ at the two-loop
level, which are independent of the applied renormalization
scheme.
   To the best of our knowledge, however, no complete two-loop calculation
in a manifestly Lorentz-invariant formulation of baryon ChPT
(BChPT) has been performed so far.
   The determination of the nucleon mass up to a given order is one
of the simplest calculations that can be performed in BChPT up to
that order.
   This makes it the ideal physical quantity to
perform a complete and consistent calculation at the two-loop
level.
   In this Letter we present the results of
such a calculation up to and including order ${\cal O}(q^6)$ using
the reformulated infrared renormalization.

\section{Lagrangian and power counting}

   The effective Lagrangian relevant for the calculation of the
nucleon mass to order ${\cal O}(q^6)$ is given by the sum of a
purely mesonic and a one-nucleon part,
   \begin{equation}\label{Lag}
    {\cal L}_{\rm eff} = {\cal L}_{2} + {\cal L}_{4} + {\cal L}^{(1)}_{{\pi}N} +
{\cal L}^{(2)}_{{\pi}N} + {\cal L}^{(3)}_{{\pi}N} + {\cal
L}^{(4)}_{{\pi}N} + {\cal L}^{(5)}_{{\pi}N} + {\cal
L}^{(6)}_{{\pi}N} + \cdots.
\end{equation}
   The purely mesonic Lagrangian at order ${\cal O}(q^2)$ is given
in Ref.~\cite{Gasser:1983yg}.
   Reference \cite{Gasser:1987rb} contains
the mesonic Lagrangian at order ${\cal O}(q^4)$ as well as the
lowest-order and next-to-leading-order nucleonic Lagrangians.
   The Lagrangians of the nucleon sector at order ${\cal O}(q^3)$
and ${\cal O}(q^4)$ can be found in
Refs.~\cite{Ecker:1995rk,Fettes:2000gb}.
   The complete Lagrangians at order ${\cal O}(q^5)$and ${\cal O}(q^6)$ have
not yet been constructed.
   However, up to the order we are considering, the nucleon mass
does not receive any contributions from the Lagrangian at order
${\cal O}(q^5)$, since only even powers of the pion mass can be
generated from contact terms.
   The contributions from the Lagrangian at order ${\cal O}(q^6)$
will be of the form $\hat{g}_1 M^6$, where $\hat{g}_1$ denotes a
linear combination of low-energy coupling constants (LECs) from
${\cal L}^{(6)}_{{\pi}N}$.

   We use the following standard power counting
\cite{Weinberg:1991um,Ecker:1994gg}:
   Each loop integration in $n$
dimensions is counted as $q^n$, a pion propagator as $q^{-2}$, a
nucleon propagator as $q^{-1}$ and vertices derived from ${\cal
L}_{i}$ and ${\cal L}^{(j)}_{{\pi}N}$ as $q^i$ and $q^j$,
respectively.

\section{Results and discussion}

\begin{figure}
\begin{center}
\epsfig{file=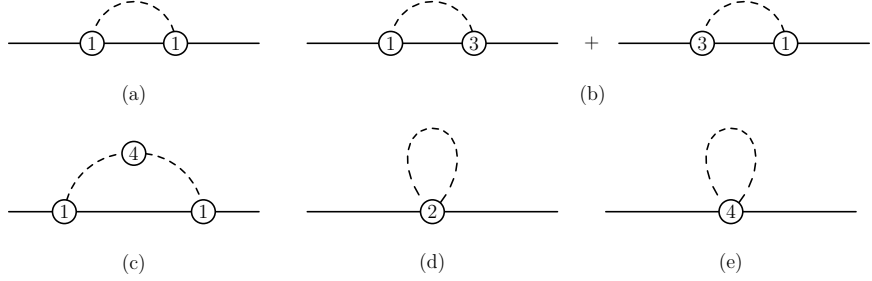,width=0.7\textwidth}
\end{center}
\caption{One-loop diagrams contributing to the nucleon mass up to
order ${\cal O}(q^6)$.\label{MassOneLoopDia}}
\end{figure}

\begin{figure}
\begin{center}
\epsfig{file=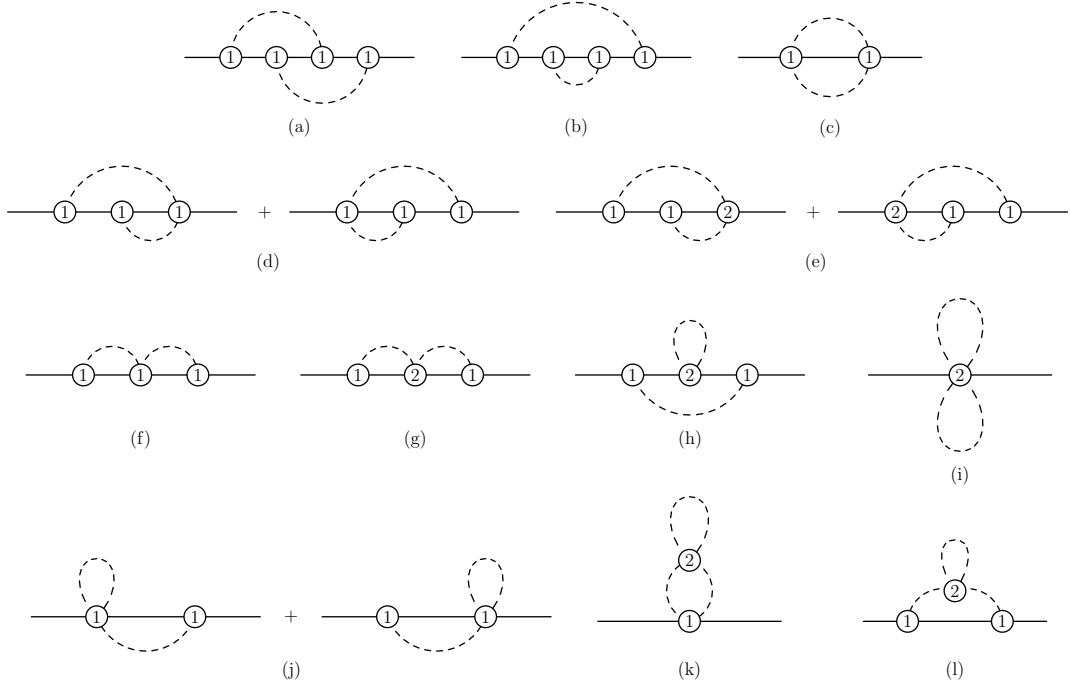,width=0.55\textwidth,angle=-90}
\end{center}
\caption{Two-loop diagrams contributing to the nucleon mass up to
order ${\cal O}(q^6)$.\label{MassTwoLoopDia}}
\end{figure}

   Figure~\ref{MassOneLoopDia} shows the one-loop diagrams which
generate nonvanishing contributions to the nucleon mass up to
order ${\cal O}(q^6)$.
   Diagrams with mass insertions in the nucleon propagator are
taken into account by shifting the mass in the undressed nucleon
propagator \cite{Becher:1999he}.
   Diagrams (a) and (d) are of order ${\cal O}(q^3)$ and ${\cal
O}(q^4)$, respectively.
   Diagrams (b) and (c) are of order ${\cal O}(q^5)$, while diagram (e)
counts as ${\cal O}(q^6)$.
  The relevant two-loop diagrams are shown in
Fig.~\ref{MassTwoLoopDia}.
   According to the power counting there are additional diagrams
at the given order, which however give vanishing contributions to
the nucleon mass up to this order.
   An example would be diagram \ref{MassTwoLoopDia}~(c) with one first-order vertex replaced by a
second-order one.

   The renormalization of one- and two-loop integrals is performed
using the reformulated infrared renormalization of
Ref.~\cite{Schindler:2003xv}.
   Details of the renormalization procedure will be given in a
forthcoming publication \cite{InPrep}.

   Since the perturbative expansion is performed around a ground state
realized in the Nambu-Goldstone mode, the quark mass expansion of
physical quantities contains analytic as well as nonanalytic terms
\cite{Li:1971vr}.
   The chiral expansion of the
nucleon mass up to order ${\cal O}(q^6)$ reads
\begin{eqnarray}\label{MassExp}
    m_N &=& m +k_1 M^2 +k_2 \,M^3 +k_3 M^4 \logM
+ k_4 M^4  + k_5 M^5\logM + k_6 M^5  \nonumber\\
    && + k_7 M^6 \logMsq + k_8 M^6 \logM + k_9 M^6 \,,
\end{eqnarray}
where $M^2=2B\hat{m}$ is the leading-order expression for the pion
mass squared in terms of the average light quark mass,
$\hat{m}=(m_u+m_d)/2$ \cite{Gasser:1983yg}, and $m$ stands for the
nucleon mass in the chiral limit.
   Combining the contributions from contact interactions with the
one- and two-loop results the coefficients $k_i$ are given by
\begin{eqnarray}\label{MassCoeff}
   k_1 &=& -4 c_1 \,,\nonumber\\
   k_2 &=& -\frac{3 g_A^2}{32 \pi F^2} \,,\nonumber\\
   k_3 &=& - \frac{3}{32\pi^2F^2m}
\left(g_A^2-8c_1m +c_2 m +4c_3m\right)\,,\nonumber\\
   k_4 &=& - \hat{e}_1
-\frac{3}{128\pi^2F^2m}\left(2 g_A^2-c_2m\right) \,,\nonumber\\
   k_5 &=& \frac{3 g_A^2}{1024\pi^3 F^4}\,\left(16g_A^2-3\right) \,,\nonumber\\
   k_6 &=& \frac{3 g_A^2}{256 \pi^3 F^4}\, \left[ g_A^2 + \frac{\pi^2 F^2}{m^2}
-8\pi^2(3l_3-2l_4) -\frac{32\pi^2F^2}{g_A}\,(2d_{16}-d_{18}) \right] \,,\nonumber\\
   k_7 &=&  -\frac{3}{256\pi^4 F^4 m}\,\left[g_A^2 - 6 c_1 m + c_2m +4c_3m
  \right]  \,,\nonumber\\
   k_8 &=& -\frac{g_A^4}{64\pi^4 F^4 m}-\frac{g_A^2}{1024\pi^4 F^4 m^2}\left[384 \pi^2 F^2 c_1
+5m+192\pi^2m(2l_3-l_4)\right] \nonumber\\
   && -\frac{3g_A}{8\pi^2F^2m}\left[2d_{16}-d_{18}\right] +\frac{3}{256\pi^4F^4}\left[ 2c_1-c_3\right]
\nonumber\\
   &&+\frac{1}{8\pi^2F^2m}\left[ 6c_1c_2
-12\hat{e}_2 m-6\hat{e}_3m -e_{16}m  \right]\,,\nonumber\\
   k_9 &=&  \hat{g}_1 - \frac{g_A^4}{24576\pi^4 F^4 m}\left( 49+288\pi^2 \right)
-\frac{3g_A}{16\pi^2 F^2 m} \left( 2d_{16}-d_{18} \right)
\nonumber\\
   && -\frac{g_A^2}{1536\pi^4 F^4 m^3}\left[ m^2(1+18\pi^2)-12\pi^2 F^2 +144\pi^2 m^2\left(3l_3-l_4\right)
\right.  \nonumber\\
   && +\left. 288\pi^2 F^2 m c_1 -24\pi^2 m^3 \left(c_3-2c_4\right)\right] \nonumber\\
   && +\frac{1}{6144\pi^4 F^4 m}\left[3-1152\pi^2 F^2 c_1 c_2 +1152\pi^2 F^2 m\, \hat{e}_3  +320\pi^2F^2m\, e_{16}\right]\,.
\end{eqnarray}
   To simplify the notation we use
\begin{eqnarray*}
  \hat{e}_1 &=& 16e_{38}+2e_{115}+2e_{116} \,,\\
  \hat{e}_2 &=& 2e_{14}+2e_{19}-e_{36}-4e_{38} \,, \\
  \hat{e}_3 &=& e_{15}+e_{20}+e_{35}
\end{eqnarray*}
for combinations of fourth-order baryonic LECs, while $\hat{g}_1$
denotes a combination of LECs from the Lagrangian at order ${\cal
O}(q^6)$.

   In general, the expressions of the coefficients in the chiral
expansion of a physical quantity differ in various renormalization
schemes, since analytic contributions can be absorbed by
redefining LECs.
   However, this is not possible for the leading nonanalytic
terms, which therefore have to agree in all renormalization
schemes.
   Comparing our result with the HBChPT calculation of
\cite{McGovern:1998tm}, we see that the expressions for the
coefficients $k_2$, $k_3$, and $k_5$ agree as expected.
   At order ${\cal O}(q^6)$ also the coefficient $k_7$ has to be
the same in all renormalization schemes.
   Note that, while $k_6 M^5$ and $k_8 M^6 \logM$ are
nonanalytic in the quark masses, the algebraic form of the
coefficients $k_6$ and $k_8$ are renormalization scheme
\emph{dependent}.
   This is due to the different treatment of one-loop diagrams.
   The counterterms for one-loop subdiagrams depend on the
renormalization scheme and produce nonanalytic terms proportional
to $M^5$ and $M^6 \logM$ when used as vertices in counterterm
diagrams.

   The numerical contributions from higher-order terms cannot be
determined so far, since most expressions in Eq.~(\ref{MassCoeff})
contain unknown LECs from the Lagrangians of order ${\cal O}(q^4)$
and higher.
   Only the coefficient $k_5$ is free of these higher-order LECs
and is given in terms of the axial-vector coupling constant $g_A$
and $F$.
   While the values for both $g_A$ and $F$ should be taken in the chiral
limit, we evaluate $k_5$ using the physical values
$g_A=1.2695(29)$ \cite{Yao:2006px} and $F_\pi=92.42(26)$ MeV in
order to get an estimate of higher-order corrections.
   Setting $\mu=m_N$, $m_N=(m_p+m_n)/2=938.92$ MeV, and $M=M_{\pi^+}=139.57$ MeV
we obtain $k_5 M^5 \ln(M/m_N) = -3.8$ MeV.
   This amounts to approximately $25$\% of the leading nonanalytic contribution at one-loop
order, $k_2 M^3$.
   For a discussion of the importance of the terms at order ${\cal
O}(q^5)$ at unphysical quark masses as used in lattice
extrapolations see Ref.~\cite{McGovern:2006fm}.

\section{Summary}
   Using the reformulated infrared regularization \cite{Schindler:2003xv} we have calculated
the nucleon mass up to and including order ${\cal O}(q^6)$.
   This is the first complete two-loop calculation in manifestly
Lorentz-invariant baryon chiral perturbation theory.
   The applied renormalization scheme preserves the standard power counting
and respects all symmetries.
   Our results for the renormalization scheme independent terms agree
with the HBChPT results of Ref.~\cite{McGovern:1998tm}.

\begin{acknowledgments}
The work of M.~R.~Schindler, D.~Djukanovic, and J.~Gegelia  was
supported by the Deutsche Forschungsgemeinschaft (SFB 443 and SCHE
459/2-1).

\end{acknowledgments}

\end{document}